\documentclass{article}

\usepackage{PRIMEarxiv}

\usepackage[utf8]{inputenc} 
\usepackage[T1]{fontenc}    
\usepackage{hyperref}       
\usepackage{url}            
\usepackage{booktabs}       
\usepackage{amsfonts}       
\usepackage{nicefrac}       
\usepackage{microtype}      
\usepackage{lipsum}
\usepackage{fancyhdr}       
\usepackage{graphicx}       
\graphicspath{{media/}}     

\title{The Ethics of Interactions: Mitigating Security
Threats in LLMs}

\author{
  Ashutosh Kumar, Shiv Vignesh Murthy, Sagarika Singh, Swathy Ragupathy \\
  Rochester Insititute of Technology \\
  Rochester, NY-14623, USA \\
  \texttt{\{ak1825, sm2678, ss3028, sr7788\}@rit.edu}
}

\begin{document}
\maketitle

\begin{abstract}
This paper comprehensively explores the ethical challenges arising from security threats to Large Language Models (LLMs). These intricate digital repositories are increasingly integrated into our daily lives, making them prime targets for attacks that can compromise their training data and the confidentiality of their data sources. The paper delves into the nuanced ethical repercussions of such security threats on society and individual privacy. We scrutinize five major threats—prompt injection, jailbreaking, Personal Identifiable Information (PII) exposure, sexually explicit content, and hate-based content—going beyond mere identification to assess their critical ethical consequences and the urgency they create for robust defensive strategies. The escalating reliance on LLMs underscores the crucial need for ensuring these systems operate within the bounds of ethical norms, particularly as their misuse can lead to significant societal and individual harm. We propose conceptualizing and developing an evaluative tool tailored for LLMs, which would serve a dual purpose: guiding developers and designers in preemptive fortification of backend systems and scrutinizing the ethical dimensions of LLM chatbot responses during the testing phase. By comparing LLM responses with those expected from humans in a moral context, we aim to discern the degree to which AI behaviors align with the ethical values held by a broader society. Ultimately, this paper not only underscores the ethical troubles presented by LLMs; it also highlights a path toward cultivating trust in these systems.
\end{abstract}

\section{Introduction}
\subsection{Understanding LLMs}
Large Language Models (LLMs) \cite{chang2024survey} are deep learning-based models, designed to process and generate text at scale, leveraging advanced neural network architectures such as transformers. A few crucial processes are involved in how LLMs operate. Initially, a sizable corpus of text data, including books, journals, and web pages, is used to train the model, which involves exposing the model to a wide range of language tasks, enabling it to learn representations of words, phrases, and sentences in a way that captures their meanings and relationships. The model is trained using deep learning methods like neural networks, with several processing layers and an encoder-decoder architecture. They comprise millions or even billions of parameters, which enable them to capture complex linguistic patterns and semantic relationships. After training, the model can produce contextually relevant and coherent text responding to the input text.\\[5pt]
LLMs can produce coherent and fluent content even when given incomplete or unclear input by employing the language modeling technique, which involves determining the probability of each word in a sentence based on the ones that came before it. Modern LLMs are capable of zero-shot and few-shot learning, meaning they can perform tasks without specific training. This allows them to generalize to new domains and tasks with minimal additional training. LLMs, such as GPT-4 by OpenAI, Claude, Llamma by Meta AI, Gemma by Google, and Phi by Microsoft, have demonstrated advanced language understanding and generation capabilities, leading to widespread adoption in diverse applications, including natural language understanding, text generation, code generation, information retrieval, conversational AI, and more.\\[5pt]
\begin{table*}[]
    \centering
    \begin{tabular}{p{6cm}p{9.5cm}}
       \textbf{Application}  & \textbf{Description} \\ \hline \\
      \textbf{Natural Language Processing (NLP)}   & LLMs have revolutionized NLP tasks, including text classification, named entity recognition, sentiment analysis, and language translation. Their ability to understand and generate human-like text has enhanced the accuracy and performance of NLP applications.\\[5pt]
      \textbf{Conversational AI and Chatbots} & LLMs are the foundation for developing sophisticated conversational AI systems and chatbots. They enable more natural and contextually relevant interactions, improving user experiences in customer support, virtual assistants, and dialogue systems.\\[5pt]
      \textbf{Information Retrieval and Question Answering} & LLMs excel in information retrieval and question-answering tasks by comprehensively understanding and processing complex queries. They power search engines, recommendation systems, and knowledge bases, enabling users to access relevant information efficiently.\\[5pt]
      \textbf{Content Generation and Summarization} & LLMs are utilized for content generation, including automatic writing, summarization, and paraphrasing. They can produce coherent and contextually relevant text, contributing to applications such as content creation, news summarization, and document generation.\\[5pt]
      \textbf{Code Generation and Programming Assistance} & LLMs have been integrated into code generation, completion, and programming assistance tools. They aid developers in writing code, debugging, and understanding programming languages, enhancing productivity and software development processes.\\[5pt]
      \textbf{Personalization and Adaptive Systems} & LLMs enable personalized content recommendations, adaptive interfaces, and tailored user experiences. They analyze user input and behavior to deliver customized responses and services in content platforms and e-commerce applications. They are also employed in the education sector, where tools are being developed to produce instructional content and give students personalized feedback. Customer care chatbots that can converse with clients in natural language and offer tailored responses are being created using LLMs.\\[5pt] \hline
    \end{tabular}
    \caption{A detailed look at LLMs and their integration into contemporary applications}
    \label{tab:llm_application}
\end{table*}
Although LLMs have demonstrated remarkable performance and a broad range of application possibilities as shown in Table \ref{tab:llm_application}, there are several possible risks involved in using LLMs. One problem is that social biases in the training data may be inherited and amplified by LLMs, which could result in unethical and unfair outcomes for the model. Another risk is that LLMs may, on purpose or in response to particular cues, produce improper, deceptive, or harmful content. This may have profound effects on people and society at large. The complexity and scale of LLMs have increased privacy concerns, especially about data sharing and potential exploitation. More questions are raised if models are made public after being trained on private data. LLMs frequently commit terms from their training sets to memory, which could be used by a malicious party to obtain confidential or personal information and jeopardize individual privacy.\\[5pt]
Understanding LLMs involves recognizing their capabilities in language processing, their potential impact on diverse applications, and the need to address challenges related to responsible use and ethical considerations. As LLMs evolve, ongoing research and development efforts are essential to harnessing their potential while mitigating risks and ensuring their alignment with societal values.
\subsection{Identifying Vulnerabilities in LLMs}
\subsubsection{Prompt Injection}
A complex manipulation of input for Large Language Models (LLMs) is a critical concern within the realm of AI ethics. Manipulating the input by injecting biased, false, or harmful prompts into these models leads to outputs that perpetuate bias, encourage destructive behavior, or distort or leak sensitive information. Vulnerabilities stemming from prompt injection evoke ethical concerns that warrant proactive measures. \\[5pt]
Multiple avenues of LLMs are vulnerable to prompt injection:
\begin{itemize}
     \item \textbf{Data Poisoning}: The integrity of LLMs relies heavily on the nature of their training data. Prompt manipulation during fine-tuning can introduce biases, leading to unintended discrimination in real time. The consequences amplify when these models undertake content moderation roles, potentially perpetuating stereotypes against specific ethnic groups. 
    \item \textbf{Model Inversion Attacks}: Proprietary and powerful language and learning models (LLMs) like ChatGPT operate with undisclosed architectural details, withholding information from the general public. Attempts to reverse engineer such models pose significant ethical concerns, as adversaries could reconstruct a potent replica of the original model. This replicated version might be exploited for malicious purposes, effectively becoming an 'evil twin' of the authentic model. This scenario highlights the ethical dilemma surrounding the transparency of LLMs. Adversaries might craft prompts designed to elicit specific responses that indirectly reveal details about the model's architecture.
    \item \textbf{Bypassing Guardrails}: Sophisticated LLMs, such as ChatGPT, are equipped with guardrails that detect prompts or inputs suggesting the creation of fictional identities or personas. Detection algorithms could flag or restrict outputs that heavily focus on generating detailed personal information about non-existent individuals. Fabricating false personas could lead to misinformation, deceit, or the potential for malicious activities, contradicting ethical usage. However, guardrails are only partially secure and carry some vulnerabilities that adversaries hunt to exploit. Adversaries might gradually introduce fictional character details into conversations or prompts to bypass detection algorithms. Starting with seemingly innocent queries before transitioning progressively to more elaborate fictional details could evade detection. Employing coded language, allusions, or indirect references to describe fabricated personas makes it harder for detection algorithms to flag such inputs.
\end{itemize}
\subsubsection{Jailbreaking LLMs}
In the context of LLMs, jailbreaking \cite{zhuo2023red} aims to circumvent the restrictions imposed by the model owner or the hosting platform to achieve unauthorized access to the Language Models' internal functionalities and protocols. Unrestricted access through jailbreaking jeopardizes the model's integrity. Attackers could exploit this to infiltrate and manipulate critical algorithms or datasets, potentially leading to distortions, biases, or unauthorized output alterations. Furthermore, Jailbreaking may pave the way for injecting malicious code, compromising the system’s integrity, and perpetuating biases.
\subsubsection{Personal Identity Information (PII) Leaks}
LLMs are trained on vast volumes of data encompassing information from the World Wide Web. These datasets contain confidential and sensitive information, which makes these models vulnerable to leakage. Several reasons make LLMs susceptible to PII leakage. Although overfitting and memorization are common reasons, these can be mitigated relatively easily by tuning the model to rely less on memorization and focus more on generalizability. Attackers leverage prompts to analyze LLM behavior, uncovering sensitive informational tendencies. Real-world prompts tied to recent events can elicit context-associated data. Patterns in model responses might inadvertently reveal personal identifiers like names, phone numbers, SSNs, or financial data like credit card numbers. Exploiting these leaks can lead to identity theft, financial fraud, and severe repercussions for affected individuals or organizations.
\subsubsection{Sexual and Hateful Content}
Data Poisoning and Bypassing Guardrails may provoke LLMs to generate controversial information, such as sexual or hateful content, which presents several ethical and moral concerns. LLMs excel at developing natural language that closely resembles the style of humans. These models can generate responses that flow naturally in conversations, exhibiting nuances, humor, and contextually appropriate replies. This realism contributes to blurring the distinction between machine-generated and human-generated content.\\[5pt]
LLMs capable of generating sexually explicit content might influence unrealistic expectations about relationships or intimacy, potentially manipulating young minds. Inappropriate outputs from LLMs might reinforce derogatory remarks or stereotypes about genders, contributing to the perpetuation of societal biases and discrimination. Misinformation or misguided advice generated by LLMs regarding sexual health or practices could lead to unsafe behaviors among adolescents who lack comprehensive sex education. Exposure to unrealistic or inappropriate content might impact children and adolescents' emotional and psychological well-being, shaping their perceptions and behaviors in unhealthy ways. Reinforcing stereotypes or derogatory remarks through LLM outputs can contribute to normalizing discriminatory attitudes or behaviors among younger audiences.\\[5pt]
Elderly individuals, influenced by societal changes, might hold onto outdated opinions regarding race, gender, or social norms. LLMs might inadvertently validate or reinforce these obsolete perspectives. LLMs might reinforce resistance to adopting new societal norms or technological advancements by generating outputs aligned with traditional, possibly outdated, viewpoints. If LLM-generated content aligns with outdated biases, it could reinforce discriminatory attitudes or hinder the acceptance of more inclusive and progressive societal norms. Elderly individuals relying on LLM-generated content might face social isolation or be susceptible to misinformation if exposed primarily to content reinforcing outdated opinions.
\section{Why Ethics Matter in LLM Attacks?}
A critical factor in the creation and use of LLMs is ethics. Because LLMs can produce information that might be interpreted positively or negatively, it is essential to have proactive ethical frameworks and legislative procedures to regulate their proper usage and hold them accountable for the results. Important ethical factors in LLMs include interpretability and explainability. Understanding LLMs' decision-making processes is difficult due to their "black-box" nature, essential for gaining public acceptance and trust—especially in delicate areas. Their efficacy and reliability are restricted by their lack of operational understanding, even with their sophisticated skills.\\[5pt]
The rapid advancement and widespread adoption of LLMs make their potential compromise through malicious attacks an urgent ethical concern. Attacks aim to deliberately manipulate LLM responses to spread misinformation, bias, hate speech, or inappropriate content that could significantly impact public discourse and decision-making. It is, therefore, critical to safeguard LLMs based on ethical norms. LLMs are being integrated into sensitive domains like healthcare, education, law, and policymaking, where reliability and truthfulness are paramount. Compromised models that generate convincing, misleading, or biased claims could undermine evidence-based decision-making and erode public trust. The diffusion of toxic, discriminatory, and unreliable content threatens ethical values like wisdom, dignity, equality, and social cohesion that underpin a just society. Furthermore, attacks to expose private data from an LLM's training set or breach its secure systems raise ethical issues around consent, privacy, identity theft, and surveillance. Such violations contravene ethical duties to respect individual autonomy and prevent harm. On a broader level, manipulated LLMs that falsely portray minorities or marginalized groups risk reinforcing structural oppression.\\[5pt]
For instance, employees at Samsung Semiconductor \cite{business2023samsung} accidentally leaked company secrets through ChatGPT prompts, according to a report by Business Today. In response, Samsung's chief has warned staff against repeating such mistakes, threatening to block access to ChatGPT on the company network if it happens again. This incident highlights the risks of sharing sensitive data with large language models (LLMs) like ChatGPT \cite{derner2023beyond}. Even though OpenAI claims to remove personally identifiable information, it could be retained and reproduced once confidential data is submitted to these models. Samsung Semiconductor is developing an internal AI assistant to avoid further data leaks. However, it will have strict data constraints, only processing short prompts under 1,024 bytes (Business Today, 2023). LLMs are trained on vast datasets, so leaking proprietary information makes it available to other users. Over time, this could empower competitors with valuable insider knowledge.  This demonstrates how organizations must weigh the benefits of AI tools against data security risks through governance policies and access controls. The Samsung case underlines the need for employees to be cautious when sharing confidential data with public chatbot systems despite their conversational convenience.\\[5pt]
Therefore, Safeguarding LLMs' integrity is an ethical obligation to foster their purposeful development aligned with social goods. Techniques like transparency requirements, controlled testing for emergent harms, and instituting recourse mechanisms can help continually assess and address threats from attacks. Conceptualizing and governing LLM safety through the lens of ethics provides the imperative and moral basis to motivate corporate accountability and remedy issues as they emerge. Establishing such ethical foundations is indispensable to building public trust in LLMs' productive integration into society.
\section{Potential Misuse and Security Concerns}
Examining ethical principles for Large Language Models (LLMs) \cite{10255221} is a crucial aspect of the evolution of AI technology. It involves scrutinizing how these models handle sensitive topics, maintain privacy, and avoid biases, ensuring they align with human values and societal norms. This scrutiny guides the responsible development and deployment of LLMs and shapes public trust and acceptance, balancing technological advancement with ethical responsibility. As LLMs become more integral to daily life, their ethical framework becomes more significant in defining their role and impact on society.Table \ref{tab:llm_ethics} describes the ethical concerns considering LLMs.\\[5pt]
\begin{table*}[]
\centering
\begin{tabular}{p{0.17\textwidth}p{0.17\textwidth}p{0.17\textwidth}p{0.17\textwidth}p{0.17\textwidth}}
\textbf{Misinformation and Societal Implications} & \textbf{Identity Theft from Training Data} & \textbf{Bias Amplification} & \textbf{Economic Repercussions} & \textbf{Privacy Challenges}\\ \hline \\
LLM-generated misinformation threatens evidence-based decision-making on critical issues like climate change and public health. & PII extracted from training data enables digital impersonation and phishing, violating individual autonomy. & Biased training data and targeted prompts can amplify discrimination against groups with less oversight power. & LLM manipulation in key sectors risks eroding credibility, value generation, and public trust in system outputs. & Ability to reconstruct images and match identities from model outputs threatens privacy rights and consent.\\ \hline \\
Disinformation campaigns powered by LLMs risk skewing elections and eroding shared foundations of truth. & Incomplete anonymization allows tracing data to original, non-consenting authors. & Restorative steps complicated by power imbalances; consequences entrench demographic inequalities. & Financial impacts may disproportionately affect vulnerable communities while obscuring accountability. & Lack of data sourcing transparency hinders ethical review and risks exposing private conversations.\\ \hline
\end{tabular}
\caption{Ethical Concerns in Large Language Models}
\label{tab:llm_ethics}
\end{table*}
\section{Towards Ethical Mitigation: A Proposed Methodology}
The tool is designed to enhance the guidelines and strategies for securing large language model (LLM) systems. It hypothesizes a robust solution for identifying and mitigating unethical or harmful user interactions.
\subsection{User Prompt Reception} The system initially receives the user's prompt through the LLM user interface.
\subsection{Prompt Classification Engine}
\begin{itemize}
    \item \textbf{Prior Analysis}: The prompt undergoes an initial analysis where it is scanned for characteristics of these five categories of vulnerabilities: prompt injection attack, jailbreak attempt, personally identifiable information (PII), sexual content, and hateful content.
    \item \textbf{Probability or Likelihood Calculation}: For each category, the system calculates the likelihood of the prompt belonging to each category as [p1, p2, p3, p4, p5] using respective classification methods, like Natural Language Processing (NLP) for sexual and hateful prompts. The probabilities are not necessarily an exact emulation of the nature of the prompt because a significant number of attacks are based on a sequence of prompts designed to break the LLM’s application structure.
    \item \textbf{Primary Category Identification}: The system identifies the primary attack category and launches the response design phase based on the highest probability beyond a certain threshold. However, multiple types of attacks can be carried out simultaneously using a combined response design because the ultimate purpose of this tool is to build a shield against all those attacks.
\end{itemize}

\begin{figure}
    \centering
    \includegraphics[width=\linewidth]{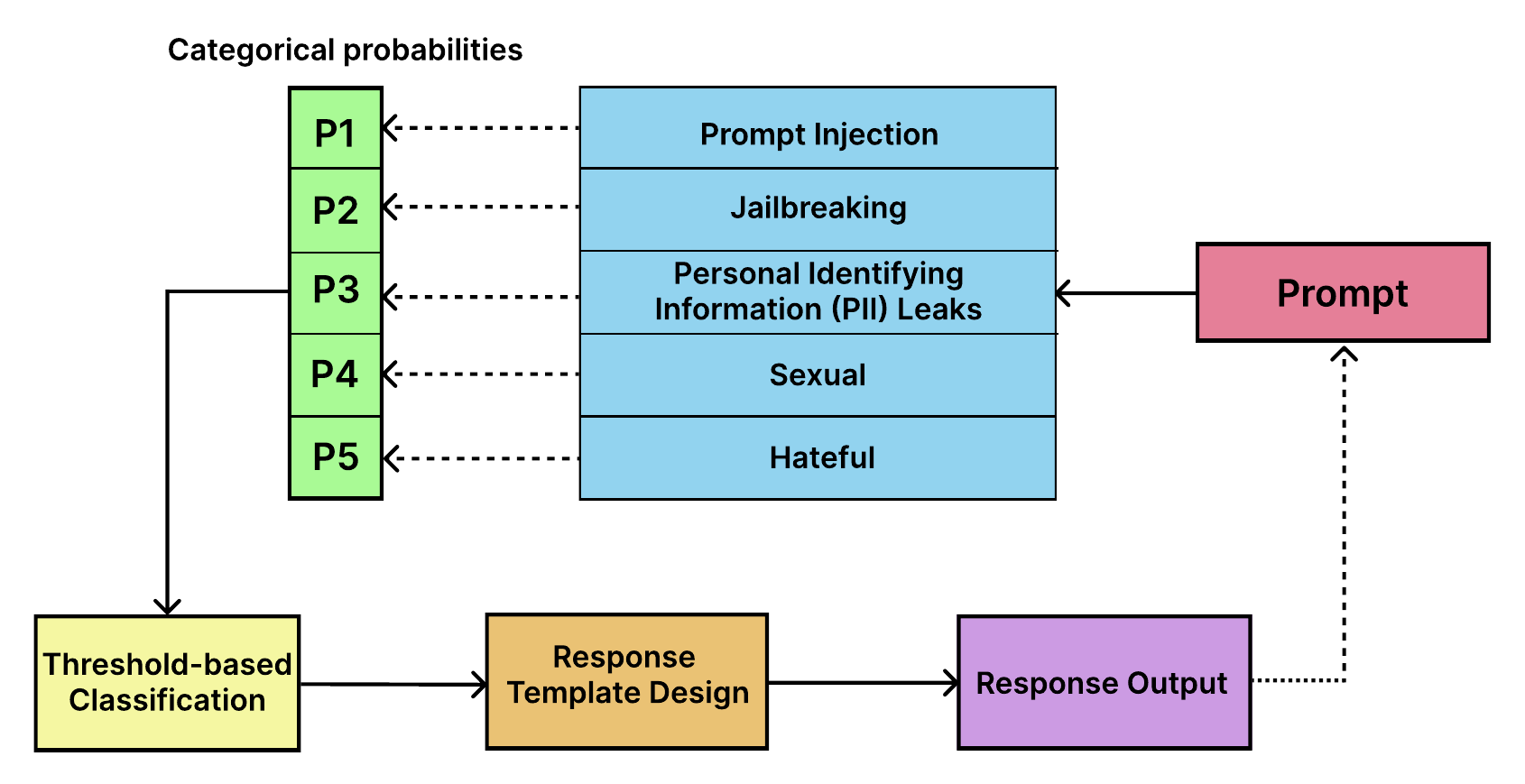}
    \caption{A tool designed to mitigate LLM security risks}
    \label{fig:enter-label}
\end{figure}

\subsection{Ethical and Security Compliance Check}
\begin{itemize}
    \item Based on the identified primary category, the system checks specific compliance rules and ethical guidelines (if already in place) relevant to that category, or if the LLM system is still in the testing phase, the respective ethical policies can be designed based on risk analysis using this tool and with the help of stakeholders including ethicists, designers, and developers.
    \item This ensures that the response is tailored to mitigate the potential risks associated with the identified category. The LLM system must respond sensitively to the incoming prompt so as not to trigger more enhanced attacks on its application database and system architecture.
\end{itemize}
\subsection{Response Design Phase}
\begin{itemize}
    \item \textbf{Template Selection}: A response template is chosen depending on the primary category. These templates are pre-designed to handle specific categories of attacks. Designing a template for a combined attack is tricky, as there might be hidden repercussions with the user interactions. Designers and developers must sit with ethical scientists, sociologists, and psychology experts to formulate the designs and trigger keywords and thresholds for each template.
    \item \textbf{Customization and Filtering}: The response is customized to the specific prompt, ensuring no unethical or harmful content is included. This may involve filtering out sensitive information or reframing the answer to avoid endorsing or propagating harmful content. The thing to note here is that these are response design templates, not the actual responses.
\end{itemize}
\subsection{Response Delivery}
The ethically compliant and secure response is delivered to the user.
\subsection{Monitoring and Feedback Loop}
\begin{itemize}
    \item The system continuously monitors its performance and the accuracy of its classifications, probably with the help of manual oversight in its initial design and implementation phase.
    \item Feedback from these monitoring processes is used to refine the classification algorithms and response templates.
\end{itemize}
\subsection{Applications}
\begin{itemize}
    \item Testing for ethical interactions at beta-level (test phase) LLM software and further designing ethical guidelines
    \item Integrating into existing LLM chatbots to monitor incoming prompts and update the security protocols to ensure ethical interaction between the user and the LLM
    \item Being open-source, it can rely on constant updation of ethical and security bugs and create a feedback loop for AI ethical security in general
\end{itemize}

\subsection{Challenges}
\subsubsection{Contextual Assessment}
Recognizing that a single prompt might fit within a specific attack category is essential, but this is only sometimes the case. Such a prompt is often merely a fragment of a broader conversation, especially noticeable at the start of an interaction. The challenge lies in determining the optimal point for activating the prompt classification mechanism. One approach to consider is using a dictionary, which maps the sequences of user prompts to potential attack categories. This would function like assigning a preliminary probability of an attack based on key-value pairs accumulated during testing phases.
\subsubsection{Shielded System Design}
Safeguard the tool by concealing its inner workings, ensuring it remains secure. Implement separate, isolated components within the large language model (LLM) backend and the tool interface. This separation prevents external reprogramming and loss of control over the tool. Remember, increasing the number of features can also increase potential vulnerabilities in the system.
\subsubsection{Auto-Disable Functionality}
If a node in the tool malfunctions, it should automatically deactivate, allowing the LLM to revert to its original, pre-tool state with updated protocols. This modular design ensures the tool can be seamlessly detached from the LLM system when necessary.
\section{Preemptive Ethical Measures}
Integrating your tool into designing and developing Large Language Models (LLMs) can ensure ethical compliance, data integrity, user-friendly design, and robust security. 

Some pre-emptive ethical measures organizations could take are:
\subsection{Ethical Compliance and Transparency}
\begin{itemize}
    \item \textbf{Ethical Guideline Development}: Establish clear ethical guidelines for LLM usage, including respect for privacy, non-discrimination, and avoiding harmful content.
    \item \textbf{Transparent Decision-Making}: Document and make the decision-making process behind the LLM's design transparent, especially regarding ethical considerations. The biases of template designers can creep into the developmental stage, so it’s essential to maintain transparency.
    \item \textbf{Stakeholder Engagement}: Involve diverse stakeholders, including ethicists, in the development process to ensure a wide range of perspectives and concerns are considered. Organizations should recruit ethical scientists to ensure the development of guidelines and the overall impact of their decisions at each step of the process.
    \item \textbf{Ethics Training for Developers}: Provide ethics training for developers and designers to sensitize them to potential ethical issues in LLM development. The entire procedure requires the tech stack and a whole set of humanities to execute this.
\end{itemize}
\subsection{User Interface Design}
\begin{itemize}
    \item \textbf{Intuitive Reporting Mechanisms}: Design user interfaces with easy-to-use reporting mechanisms for unethical or problematic content generated by the LLM.
    \item \textbf{User Consent and Control}: Implement clear consent protocols for users, letting them know how their data is used and giving them control over their interaction with the LLM.
    \item \textbf{Accessible Design}: Ensure the interface is accessible to diverse users, including those with disabilities, to promote inclusivity.
\end{itemize}
\subsection{Robust Security Protocols}
\begin{itemize}
    \item \textbf{Regular Security Audits}: Conduct regular security audits to identify and address vulnerabilities in the LLM system.
    \item \textbf{Advanced Threat Detection}: Integrate advanced threat detection systems to identify and mitigate potential security breaches or misuses preemptively.
    \item \textbf{Data Protection Measures}: Implement robust data protection measures, such as encryption and secure data storage, to safeguard user data.
\end{itemize}
\subsection{Continuous Monitoring and Evaluation}
\begin{itemize}
    \item \textbf{Real-Time Monitoring}: Use your tool to monitor LLM outputs in real-time, quickly identifying and addressing ethical issues.
    \item \textbf{Feedback Loops}: Establish feedback loops where user feedback and monitoring insights are used continuously to improve the LLM’s ethical compliance.
    \item \textbf{Impact Assessment}: Regularly assess the impact of the LLM on users and society to ensure it aligns with ethical and societal values.
\end{itemize}
\subsection{Training and Development}
\begin{itemize}
    \item \textbf{Continuous Learning}: Incorporate mechanisms for the LLM to learn from its interactions and improve its ethical decision-making capabilities.
    \item \textbf{Developer and User Education}: Educate developers and users about the ethical use of LLMs and the importance of data integrity and security.
\end{itemize}
By implementing these measures, organizations can ensure that their LLMs are not only ethically compliant but also resilient to various challenges, thereby maintaining user trust and the integrity of their systems.
\section{Ethical Response to LLM Attacks}
The "AI response spectrum" refers to the range of possible responses an artificial intelligence (AI) system can generate in response to various inputs, prompts, or queries. This spectrum encompasses the diversity of potential outputs that an AI model, such as a Large Language Model (LLM), can produce, ranging from accurate and helpful responses to biased, harmful, or incorrect outputs. Understanding the AI response spectrum is crucial for evaluating the capabilities and limitations of AI systems and addressing ethical considerations and potential risks associated with their use. By mapping human responses onto the AI response spectrum, researchers and developers can analyze how AI systems interpret and process human input and work towards ensuring that the generated responses align with ethical standards, human intentions, and societal values.

Efforts to manage and regulate the AI response spectrum involve fine-tuning AI models on instruction-formatted data, aligning models using human feedback, and promoting ethical and responsible AI usage through explainability and accountability measures (7, 30). These approaches aim to steer AI responses toward being helpful, honest, and harmless while mitigating the risks of biased, harmful, or incorrect outputs. Human reactions to the diverse AI response spectrum can vary significantly based on the nature of the AI-generated outputs. 

Here are some potential human reactions corresponding to the possible AI response spectrum:
\begin{itemize}
    \item \textbf{Accurate and informative responses}: Artificial intelligence is increasingly appreciated for its accurate and insightful responses, especially in knowledge-seeking and assistance. This positive reception underscores the growing reliance on technology for decision-making and information gathering, highlighting AI's expanding role in everyday life and professional settings.
    \item \textbf{Biased or Misleading Responses}: People often express concern and skepticism towards AI systems that produce biased or misleading information, particularly when it perpetuates stereotypes or inaccuracies. This caution highlights the importance of ethical and reliable AI practices in shaping societal perceptions and decision-making processes.
    \item \textbf{Harmful or inappropriate outputs}: Users may feel discomfort, distress, or offense from toxic or improper AI-generated content, leading to a potential loss of trust in the technology. This reaction underscores the need for responsible AI development that prioritizes user well-being and trustworthiness.
    \item \textbf{Incomplete or Incoherent Responses}: Users often experience frustration and dissatisfaction with AI systems that deliver incomplete or incoherent outputs, especially when they seek clear and comprehensive answers. It highlights the importance of advancing AI to better meet user expectations for accuracy and relevance.
    \item \textbf{Ethical and Trustworthy Responses}: Humans tend to respond positively to AI-generated content that is ethical and trustworthy, fostering a sense of trust in the technology and encouraging its responsible use. This confidence is critical to integrating AI more deeply and beneficially into various aspects of life.
    \item \textbf{Creative and Novel Outputs}: Users often appreciate and engage with AI-generated content that is creative and novel, recognizing the technology's potential for innovative and imaginative outputs. This appreciation fosters a more significant interaction and interest in AI's possibilities for creativity and originality.
\end{itemize}
Understanding human reactions to the AI response spectrum is essential for evaluating the impact of AI-generated outputs on users and society. It underscores the importance of steering AI systems toward producing ethical, accurate, and helpful responses while addressing bias, harm, and misinformation concerns. Promoting transparency, explainability, and user feedback can foster trust and align AI-generated content with human values and expectations.

\bibliographystyle{unsrt}  
\bibliography{references}

\end{document}